\documentclass[reqno,12pt,a4paper]{amsart}

\usepackage{eurosym}
\usepackage{amsfonts}
\usepackage{amssymb}
\usepackage{amsmath}
\usepackage{latexsym}
\usepackage{float,subfigure,color,subfigure}
\usepackage{graphicx}
\usepackage{amsthm}
\usepackage{verbatim}

\usepackage{a4wide}

\usepackage[dvipsnames, table]{xcolor}
\definecolor{bckg}{RGB}{20.8, 20.8, 20.8}
\definecolor{oneblue}{rgb}{0.0, 0.0, 0.85}
\definecolor{Lightblue}{RGB}{214, 214, 214}
\definecolor{bluepigment}{rgb}{0.2, 0.2, 0.6}
\definecolor{charcoal}{rgb}{0.21, 0.27, 0.31}
\definecolor{denimblue}{rgb}{0.08, 0.38, 0.74}
\definecolor{Lightgray}{rgb}{0.89, 0.89, 0.89}
\definecolor{darkgrey}{rgb}{0.273, 0.281, 0.30}
\definecolor{darkelectricblue}{rgb}{0.33, 0.41, 0.47}

\usepackage[colorlinks,
           urlcolor=oneblue,
           linkcolor=denimblue,
           citecolor=NavyBlue,
           bookmarksopen=false,
           pdfpagemode=UseNone,
           pagebackref]{hyperref}

\theoremstyle{plain}
\newtheorem{theorem}{\bf Theorem}[section]

\newtheorem{experiment}[theorem]{\bf Experiment}

\newcommand{\Sfrac}[2]{{\textstyle{\frac{#1}{#2}}}}

\newcommand{\R}{\mathbb{R}}

\DeclareMathOperator{\sech}{sech}
\DeclareMathOperator{\sgn}{sgn}

\title
[Bi-directional Whitham systems]{A comparative study of bi-directional Whitham systems}

\date{\today}

\author[Dinvay]{Evgueni Dinvay}
\author[Dutykh]{Denys Dutykh}
\author[Kalisch]{Henrik Kalisch}

\address
{
	{\texttt{evgueni.dinvay@math.uib.no}},
	{\texttt{henrik.kalisch@math.uib.no}},
	Department of Mathematics, University of Bergen,
	Postbox 7800, 5020 Bergen, Norway.
}
\address
{
	{\texttt{Denys.Dutykh@univ-smb.fr}},
	LAMA, UMR5127, CNRS - Universit\'{e} Savoie Mont Blanc, Campus Scientifique,
	73376 Le Bourget-du-Lac Cedex, France. \and Univ. Grenoble Alpes, Univ.
	Savoie Mont Blanc, CNRS, LAMA, 73000 Chamb\'ery, France
}

\begin{document}

\begin{abstract}
In 1967, Whitham proposed a simplified surface water-wave model which combined
the full linear dispersion relation of the full Euler equations with a weakly
linear approximation. The equation he postulated which is now called the 
Whitham equation has recently been extended to a system of equations allowing
for bi-directional propagation of surface waves.
A number of different two-way systems have been put forward, and even though
they are similar from a modeling point of view, these systems have very different
mathematical properties.

In the current work, we review some of the existing fully dispersive systems, 
such as found in \cite{ASMP,Claassen_Johnson,Dinvay_Dutykh,Hur_Pandey,Matsuno,MKD}. 
We use state-of-the-art numerical tools to try to understand existence and stability of solutions to
the initial-value problem associated to these systems. We also put forward
a new system which is Hamiltonian and semi-linear. The new system is shown
to perform well both with regard to approximating the full Euler system,
and with regard to well posedness properties.
\end{abstract}

\maketitle

\section{Introduction}
\setcounter{equation}{0}
%
%
Consideration is given to the two-dimensional water-wave problem
for an inviscid incompressible fluid with a free surface over
an even bottom. As this problem has not been completely resolved
mathematically, there is still interest in developing new simplified
models which yield an approximate description of the waves at the
free surface in the case when the waves have distinctive properties,
such as small amplitude or large wave period. In particular, there
is the Boussinesq scaling regime which gives a good approximate
description of long waves of small-amplitude. Recently, there
has been interest in full-dispersion model which aim to give
an exact description of ``linear'' waves while still being
weakly nonlinear, and therefore accommodating some nonlinear
processes such as wave steepening. The idea of representing
the linear dynamics exactly goes back to the work of Whitham
\cite{Wh1} who conceived the equation (now called Whitham equation)
\begin{equation}
\label{dimWhitham}
	\eta_t + g \mathcal{W} \eta_x + \Sfrac 32 \Sfrac{c_0}{H} \eta \eta_x = 0
	,
\end{equation}
where
\(
	\mathcal{W} = w(-i \partial _x)
	= {\mathcal F}^{-1} w \mathcal F
\)	
is a Fourier multiplier operator defined by the dispersive function
\begin{equation}
\label{symbol}
	w(\xi) = \sqrt{ \Sfrac{ \tanh(H \xi) }{ g \xi } }, 
\end{equation}
and $c_0 = \sqrt{gH}$ is the limiting long-wave speed, 
defined in terms of the undisturbed fluid depth $H$
and the gravitational acceleration $g$.
The Fourier transform $\mathcal{F}$ and inverse transform $\mathcal{F}^{-1}$ 
are defined in the standard way, such as for example in \cite{Wh2}.
It is clear that since the operator $\mathcal{W}$ reduces to the identity for 
very long waves ($\xi \rightarrow 0$), the Whitham equation reduces
to the inviscid Burgers equation for very long waves.

Recently, Whitham's idea has been extended to the study of systems of
evolution equation which allow for bi-directional wave propagation.
In particular, in \cite{ASMP}, Aceves-S\'{a}nchez, Minzoni and Panayotaros,
found the Whitham system
\begin{align}
\label{ASMP_sys1}
	\eta_t &=
	- H \mathcal K u_x - (\eta u)_x
	, \\
\label{ASMP_sys2}
	u_t &=
	- g \eta_x - u u_x,
\end{align}
and in \cite{MKD}, it was shown how this system arises
as a Hamiltonian system from the Zakharov-Craig-Sulem formulation
of the water-wave problem using an exponential long-wave scaling.
The operator $\mathcal{K}$ is defined by the Fourier symbol
$\frac{\tanh(H \xi)}{H \xi}$, so that we have the relation
$H \mathcal{K} = g \mathcal{W}^2$.
It can be seen that since the operator $\mathcal{K}$ reduces to the identity operator for 
very long waves ($\xi \rightarrow 0)$, this Whitham system reduces
to the classical shallow-water system for very long waves.
In the remainder of this article, we will refer to the system 
\eqref{ASMP_sys1}, \eqref{ASMP_sys2} as the ASMP system.

The system \eqref{ASMP_sys1}, \eqref{ASMP_sys2}
has been studied in a number of works. In particular, it was shown in \cite{EJC}
that it admits periodic traveling-wave solutions and features a highest cusped wave on the bifurcation branch. 
The modulational stability of its periodic traveling-wave solutions
has been investigated numerically in \cite{Claassen_Johnson},
and the system has been studied numerically in the presence of an uneven bottom in \cite{VMP}.
Moreover, it was shown in \cite{Pei_Wang} that the initial-value problem on the real line is
well posed locally-in-time for data that are strictly positive and bounded
away from zero.

On the other hand, the system
\begin{align}
\label{Hur_sys1}
	\eta_t & = - H v_x - (\eta v)_x
	, \\
\label{Hur_sys2}
	v_t & = - g \mathcal{K} \eta_x - v v_x
\end{align}
was put forward by Hur and Pandey in \cite{Hur_Pandey},
and it was shown to behave somewhat more favorably than  
\eqref{ASMP_sys1}, \eqref{ASMP_sys2} with regard to modulational
instability and local well posedness (see also \cite{Claassen_Johnson}. 
We will call this system the HP system.

In the current work, it is shown how the ASMP system \eqref{ASMP_sys1}, \eqref{ASMP_sys2}
and  the HP system \eqref{Hur_sys1}, \eqref{Hur_sys2} can be related by an asymptotic change of variables. 
Using the new variables, it is also possible to obtain a Hamiltonian system which is 
much less sensitive to instabilities than either the ASMP or HP system. We also show that the new system
yields better approximations to the full water-wave problem than
any of the other bi-directional Whitham system in use so far.
We also present two other Hamiltonian systems, the right-left system,
where dependent variables are chosen to represent wave propagating
mainly to the left or to the right, and the essentially right-going
system
For the sake of completeness, we also include the Matsuno system
in our study since it is easily obtained using the Hamiltonian theory.

%
%
\section{The Hamiltonian formalism}
\setcounter{equation}{0}
%
%
A two-dimensional water-wave problem 
with the gravity $g$ and the mean depth $H$ is under consideration.
The fluid is supposed to be inviscid and incompressible with
irrotational flow.
The unknowns are the surface elevation $\eta(x,t)$
and the velocity potential $\phi(x,z,t)$.
The fluid domain is the set
$\left\{(x,z) \in \R^2 | -H < z < \eta(x,t) \right\}$
extending to infinity in the positive and negative horizontal $x$-direction.
Liquid motion is governed by the Euler system
consisting of the Laplace's equation in this domain
\begin{equation}
\label{Euler_sys1}
	\phi_{xx} + \phi_{zz} = 0 \quad \mbox{for}
	\quad x \in \mathbb{R}, \quad -H < z < \eta(x,t)
	,
\end{equation}
the Neumann boundary condition at the flat bottom
\begin{equation}
\label{Euler_sys2}
	\phi_z = 0 \quad \mbox{at}
	\quad z = -H
	,
\end{equation}
the kinematic condition at the free surface
\begin{equation}
\label{Euler_sys3}
	\eta_t+\phi_x\eta_x-\phi_z = 0 \quad \mbox{for}
	\quad x \in \mathbb{R}, \quad z = \eta(x,t)
	,
\end{equation}
and the Bernoulli equation
\begin{equation}
\label{Euler_sys4}
	\phi_t+\frac 12 \big( \phi^2_x+\phi^2_z \big) + g \eta
	= 0 \quad \mbox{for}
	\quad x \in \mathbb{R}, \quad z = \eta(x,t)
	.
\end{equation}
The total energy of the fluid motion consists of potential and kinematic energy:
\begin{equation}
\label{total_energy}
	\mathcal H = \int _{\mathbb R} \int_0^\eta gz \, dz dx +
	\frac 12 \int _{\mathbb R} \int_{-H}^\eta |\nabla \phi|^2 \, dz dx
	.
\end{equation}

It is known that the system \eqref{Euler_sys1}-\eqref{Euler_sys4} is
equivalent to a certain Hamiltonian system.
Indeed, with the trace $\Phi(x,t) = \phi(x,\eta(x,t),t)$
of the potential at the free surface
and the Dirichlet--Neumann operator $G(\eta)$
the total energy \eqref{total_energy} takes the form
\begin{equation}
\label{Euler_Hamiltonian}
	\mathcal H  = \frac 12 \int_\R g \eta^2 dx
	+ \frac 12 \int_\R \Phi G(\eta) \Phi dx
	.
\end{equation}
We regard $\mathcal H (\eta, \Phi)$ as a functional on
a dense subspace of
$L^2( \mathbb R ) \times L^2( \mathbb R )$.
We do not wish to specify smoothness of functions $\eta$, $\Phi$
and the exact domain of the functional $\mathcal H$
at this point, but we assume
its variational derivatives lie in $L^2( \mathbb R )$.
The pair $( \eta, \Phi )$ represents the canonical variables
for the Hamiltonian functional \eqref{Euler_Hamiltonian}
with the structure map
\[
	J_{\eta, \Phi}
	=
	\begin{pmatrix}
		0 & 1
		\\
		-1 & 0
	\end{pmatrix}
\]
and so the Hamiltonian equations have the form
\begin{equation}
\label{Euler_Hamilton_system}
	\eta_t = \frac{\delta \mathcal H}{\delta \Phi} 
	, \qquad
	\Phi_t = - \frac{\delta \mathcal H}{\delta \eta}
	.
\end{equation}
This evolutionary system in $L^2( \mathbb R )$ is known
to be equivalent to the Euler system \eqref{Euler_sys1}-\eqref{Euler_sys4}.
However, it does not simplify the problem since in general
there is no explicit expression for the operator $G(\eta)$.
%
%
\section{Weakly nonlinear approximations}
\setcounter{equation}{0}
%
%
In this section several approximations to Hamiltonian
\eqref{Euler_Hamiltonian} will be presented.
Each one will give rise to a system that can be considered as an
approximate model to \eqref{Euler_Hamilton_system}.
The analysis is mainly heuristic consisting of arguments
represented in \cite{Craig_Groves, Dinvay_Dutykh}, for example.

Regarding the self-adjoint operator
$D = -i \partial _x$ in $L^2( \mathbb R )$
we assume that the Dirichlet--Neumann operator appearing in
\eqref{Euler_Hamiltonian} may be approximated by the sum
$G(\eta) = G_0 + G_1(\eta)$ where
\[
	G_0(\eta) = D\tanh(HD)
	, \qquad 
	G_1(\eta)= D\eta D - G_0 \eta G_0
	. 
\]
Such substitution should not change the Hamiltonian significantly since
the remaining terms in the truncated operator $G(\eta)$ are of at least quadratic
order in $\eta$ and its derivatives.
After integration by parts (Lemma 2.1 in \cite{Dinvay_Dutykh}) it leads to
\begin{equation}
\label{Hamiltonian_eta_Phi}
	\mathcal H  = \frac 12 \int_\R
	\left(	
		g \eta^2 + \Phi G_0 \Phi
		- \eta (D \Phi)^2 - \eta (G_0 \Phi)^2
	\right)
	dx
	.
\end{equation}
One may notice the relative advantage of this approximation immediately.
Instead of integrating the system \eqref{Euler_sys1}-\eqref{Euler_sys4},
the much simpler system \eqref{Euler_Hamilton_system} with Hamiltonian
\eqref{Hamiltonian_eta_Phi} is to be solved.

In works on the surface water-wave problem, it has been common to
use unknowns other than the potential $\Phi$.
Here, we use the variable
\(
	u = \Phi_x = \phi_x + \eta_x \phi_z
	= \phi _{\tau} \sqrt{1 + \eta_x^2},
\)
which is proportional to the velocity component of the fluid $\varphi _{\tau}$
which is tangent to the surface.
This change of variables transforms the Hamiltonian \eqref{Hamiltonian_eta_Phi}
to
\begin{equation}
\label{Hamiltonian_eta_u}
	\mathcal H  = \frac 12 \int_\R
	\left(	
		g \eta^2 + u \frac{\tanh HD}{D} u
		+ \eta u^2 + \eta (\tanh HDu)^2
	\right)
	dx
	.
\end{equation}
From now on, we will refer to the pair $(\eta, u)$ as Boussinesq variables.
Note that unlike $(\eta, \Phi)$ these new variables are not canonical.
The corresponding structure map has the form
\[
	J_{\eta,u}
	=
	\begin{pmatrix}
		0 & -\partial_x
		\\
		-\partial_x & 0
	\end{pmatrix}
\]
and the Hamiltonian system \eqref{Euler_Hamilton_system}
transforms to
\begin{equation}
\label{Hamiltonian_system_eta_u}
	\eta_t = - \partial_x \frac{\delta \mathcal H}{\delta u} 
	, \qquad
	u_t = - \partial_x \frac{\delta \mathcal H}{\delta \eta}
	.
\end{equation}
It will become clear later that it is convenient to introduce yet another
change of dependent variables.
We define
the new velocity variable $v = \mathcal Ku$,
where the transformation $\mathcal{K}$ is defined by the expression
\begin{equation}
\label{Hur_transformation}
	\mathcal K = \frac{\tanh HD}{HD},
\end{equation}
which shows that it is an invertible and bounded Fourier multiplier operator.
While the physical meaning of the new velocity variable
\(
	v = \mathcal K \partial_x \Phi = i \tanh (HD) \Phi / H
\)
is not clear, it will be shown later that it can be used
to find a new system of equations which has desirable 
mathematical properties.
In these new variables the Hamiltonian functional $\mathcal H(\eta, v)$
has the form
\begin{equation}
\label{Hamiltonian_eta_v}
	\mathcal H  = \frac 12 \int_\R
	\left(	
		g \eta^2 + Hv \mathcal K^{-1} v
		+ \eta ( \mathcal K^{-1} v )^2 + \eta (HDv)^2
	\right)
	dx
\end{equation}
with the structure map
\[
	J_{\eta,v}
	=
	\begin{pmatrix}
		0 & - \mathcal K \partial_x
		\\
		- \mathcal K  \partial_x & 0
	\end{pmatrix}
\]
and the Hamiltonian system \eqref{Euler_Hamilton_system}
transforming to
\begin{equation}
\label{Hamiltonian_system_eta_v}
	\eta_t = - \mathcal K \partial_x \frac{\delta \mathcal H}{\delta v} 
	, \qquad
	v_t = - \mathcal K \partial_x \frac{\delta \mathcal H}{\delta \eta}
	.
\end{equation}

In physical problems a question often arises if there is a way
to split waves on right- and left-going components.
One possible way of doing this splitting is to regard the linearization of
the problem given in elevation-velocity variables
and then change variables \cite{MKD}.
Namely, regard the following transformation
\begin{equation}
\label{W_variable_transformation}
	r = \frac 12 (\eta + \mathcal W  u)
	, \qquad
	s = \frac 12 (\eta - \mathcal W  u)
\end{equation}
where $\mathcal W$ is supposed to be
an invertible function of the differential operator $D$.
The inverse transformation has the form
\begin{equation}
\label{inverse_W_variable_transformation}
	\eta = r + s
	, \qquad
	u = \mathcal W^{-1} (r - s)
	. 
\end{equation}
Omitting the details provided in \cite{Dinvay_Dutykh} we notice that
to split the linearized system into two independent equations
one needs to take
\begin{equation}
\label{W_transformation}
	\mathcal W
	= \sqrt{ \frac Hg \mathcal K }
	= \sqrt{ \frac{ \tanh HD }{ gD } }
	.
\end{equation}
The new variables $r$ and $s$ correspond to
right- and left-going waves, respectively.
Returning to the nonlinear theory we want to obtain
a new Hamiltonian system with respect to unknown
functions (\ref{W_variable_transformation}).
Using the variables $r$ and $s$
and integrating by parts
puts the Hamiltonian \eqref{Hamiltonian_eta_u} into the form
\begin{equation}
\label{Hamiltonian_r_s}
	\mathcal H  = \frac 12 \int_\R
	\left(	
		2g (r^2 + s^2) + (r + s) ( \mathcal W^{-1} (r - s) )^2
		+ (r + s) (\sqrt{gG_0}(r - s))^2
	\right)
	dx,
\end{equation}
with the structure map
\[
	J_{r,s}
	=
	\begin{pmatrix}
		- \mathcal W \partial_x / 2 & 0
		\\
		0 & \mathcal W \partial_x / 2,
	\end{pmatrix}
\]
and the Hamiltonian system \eqref{Euler_Hamilton_system}
transforming to
\begin{equation}
\label{Hamiltonian_system_r_s}
	r_t = - \frac 12 \mathcal W \partial_x \frac{\delta \mathcal H}{\delta r} 
	, \qquad
	s_t = \frac 12 \mathcal W \partial_x \frac{\delta \mathcal H}{\delta s}
	.
\end{equation}

In what follows we perform a Hamiltonian perturbation analysis
based on the assumption of smallness of wave gradients.
Regard a wave-field with a characteristic non-dimensional wavelength
$\lambda = l/H$, amplitude $\alpha = a/H$ and velocity
$\beta = b / \sqrt{gH}$
where $l$, $a$ and $b$ are typical dimensional parameters.
Define the small parameter $\mu = 1 / \lambda$.
Usually $\alpha$ and $\beta$ are identified and regarded
as functions of wave-number $\mu$.
For justification of the models derived below there is no need
for this identification or concretization of the dependence
$\alpha$, $\beta$ on $\mu$.
The meaning of the scaling is of course that
$\eta = H \mathcal O(\alpha)$,
$u = \sqrt{gH} \mathcal O(\beta)$ and
$HD = - iH \partial_x = \mathcal O(\mu)$.
During our derivations, omission of higher-order terms is applied only to
the Hamiltonian expressions
\eqref{Hamiltonian_eta_u}, \eqref{Hamiltonian_eta_v}.
The main idea is that high-order dispersive effects have little effect
on the energy of the motion.
Moreover, this approach guarantees that the obtained systems are Hamiltonian. 
\subsection{Matsuno model.}
The first useful system can be obtained
if we take Hamiltonian \eqref{Hamiltonian_eta_u} as it is
and find the corresponding variational derivatives.
Taking any real-valued square integrable smooth function $h$
and using the definition
\begin{multline*}
	\int _{\mathbb R} \frac{\delta \mathcal H}{\delta u} (x)h(x) dx
	=
	\left. \frac{d}{d\tau} \right| _{\tau = 0} \mathcal H( u + \tau h, \eta )
	=
	\\
	=
	\frac 12 \int_\R
	\left(
		Hh \mathcal K u + Hu \mathcal K h
		+ 2 \eta u h + 2 \eta (\tanh HDu) \tanh HDh
	\right)
	dx
\end{multline*}
one arrives after integration by parts to
\[
	\frac{\delta \mathcal H}{\delta u}
	=
	H \mathcal K u + \eta u - \tanh{HD} (\eta \tanh{HD} u)
\]
and in the same way to
\[
	\frac{\delta \mathcal H}{\delta \eta}
	=
	g \eta + \frac 12 u^2 + \frac 12 (\tanh{HD} u)^2
	.
\]
Thus System \eqref{Hamiltonian_system_eta_u} transforms to
\begin{align}
\label{Matsuno_sys1}
	\eta_t &=
	- H \mathcal K u_x - (\eta u)_x +
	\tanh{HD} (\eta \tanh{HD} u)_x
	, \\
\label{Matsuno_sys2}
	u_t &=
	- g \eta_x - u u_x - (\tanh{HD} u) \tanh{HD} u_x
\end{align}
which appeared in \cite{LannesBonneton}, and is similar to 
the systems found in \cite{Choi} and \cite{Matsuno}.
It is not known so far if the system is well posed,
but from a modeling point of view, it is sometimes
regarded as the most exact model of all the so called
bidirectional Whitham systems. 
Even though this system conserves the Hamiltonian \eqref{Hamiltonian_eta_u},
it turns out that this system is very sensitive to aliasing
due to spatial discretization.

\subsection{ASMP model.}
Simplifying the Hamiltonian through 
and appropriate scaling such as $\alpha = O(\mu^N)$
and thus discarding the last integrand
in \eqref{Hamiltonian_eta_u}, one arrives at the system
\begin{align*}
	\eta_t &=
	- H \mathcal K u_x - (\eta u)_x
	, \\
	u_t &=
	- g \eta_x - u u_x.
\end{align*}
This is the system \eqref{ASMP_sys1}, \eqref{ASMP_sys2} mentioned in the introduction.
The corresponding Hamiltonian is
\begin{equation}
\label{ASMP_Hamiltonian}
	\mathcal H  = \frac 12 \int_\R
	\left(	
		g \eta^2 + Hu \mathcal K u
		+ \eta u^2
	\right)
	dx
	.
\end{equation}
This is also a Hamiltonian system with respect to
the same Boussinesq variables
$\eta$, $u$ in the same sense as \eqref{Hamiltonian_system_eta_u}.
This model started to attract attention after it appeared
in \cite{ASMP} and \cite{MKD}.
The local well-posedness of the system \eqref{ASMP_sys1}-\eqref{ASMP_sys2}
is proved \cite{Pei_Wang} by imposing the additional condition
$\inf \eta(x, 0) > 0$ on the initial surface elevation.
It should be remarked that this condition may mean that the system
is not useful from a physical point of view 
since all surface water wave models should have the property that the mean elevation be zero.
However strictly positive solutions, like solitons for example,
have always featured prominently in the analysis of such systems.
In a recent paper by Claassen and Johnson \cite{Claassen_Johnson}
the well-posedness for more general initial data was questioned.
In fact the authors showed numerically that the ASMP system
is probably ill-posed in $L^2 (\mathbb T)$.
However, our computations suggest to assume this is not the case  
in $L^2 (\mathbb R)$ and so that the system is probably
well-posed on the real line.
We also show that periodic discretization affects numerical
computations significantly.
\subsection{Hamiltonian version of the Hur--Pandey model.}
Regarding the Hamiltonian \eqref{Hamiltonian_eta_v} given
in the new variables defined above, one may discard the last integral in the
expression and simplify the next one staying in the same
framework of accuracy up to $\mathcal O (\mu^2 \alpha \beta^2)$.
This results in the Hamiltonian
\begin{equation}
\label{Hur_Hamiltonian}
	\mathcal H  = \frac 12 \int_\R
	\left(	
		g \eta^2 + Hv \mathcal K^{-1} v
		+ \eta v^2
	\right)
	dx
\end{equation}
with the  G\^ateaux derivatives
\[
	\frac{\delta \mathcal H}{\delta v}
	=
	H \mathcal K^{-1} v + \eta v
	,
\]
\[
	\frac{\delta \mathcal H}{\delta \eta}
	=
	g \eta + \frac 12 v^2
	.
\]
Thus for the Hamiltonian \eqref{Hur_Hamiltonian},
the system \eqref{Hamiltonian_system_eta_v} has the form
\begin{align}
\label{Hamiltonian_Hur_sys1}
	\eta_t &=
	- H v_x - \mathcal K (\eta v)_x
	, \\
\label{Hamiltonian_Hur_sys2}
	v_t &=
	- g \mathcal K \eta_x - \mathcal K (vv_x)
	.
\end{align}
To the best of our knowledge this system is completely new.
One may notice that the nonlinear part of
System \eqref{Hamiltonian_Hur_sys1}-\eqref{Hamiltonian_Hur_sys2}
contains only the bounded operator $\mathcal K \partial_x$,
which could mean that it is at least a locally
well-posed system.
Moreover we shall see later that among all bidirectional
Whitham systems this is numerically the most stable one. 

If one formally substitutes the operator $\mathcal K$
into the nonlinear part of
\eqref{Hamiltonian_Hur_sys1}-\eqref{Hamiltonian_Hur_sys2}
by unity according to the long wave approximation
$\mathcal K = 1 + \mathcal O(\mu^2)$
then one arrives at the system
\begin{align*}
	\eta_t &=
	- H v_x - (\eta v)_x
	, \\
	v_t &=
	- g \mathcal K \eta_x - vv_x
	,
\end{align*}
i.e. system \eqref{Hur_sys1}, \eqref{Hur_sys2}
which was introduced by Hur \& Pandey \cite{Hur_Pandey}.
This system does well in the sense of numerical stability
comparing with ASMP model but not as well as its Hamiltonian
relative \eqref{Hamiltonian_Hur_sys1}-\eqref{Hamiltonian_Hur_sys2}.
Unlike the system
\eqref{Hamiltonian_Hur_sys1}-\eqref{Hamiltonian_Hur_sys2}
one cannot say for certain if the Hur--Pandey system is Hamiltonian
with the same structure map as the original water-wave problem.
\subsection{Right-left waves model.}
Again simplifying the Hamiltonian \eqref{Hamiltonian_r_s}
up to $\mathcal O (\mu^2 \alpha \beta^2)$
we obtain
\begin{equation}
\label{RS_Hamiltonian}
	\mathcal H  = g \int_\R
	\left(	
		r^2 + s^2 + \frac 1{2H} (r + s)(r - s)^2
	\right)
	dx
\end{equation}
with the  G\^ateaux derivatives
\[
	\frac{\delta \mathcal H}{\delta r}
	=
	2gr + \frac g{2H} (3r + s)(r - s)
	,
\]
\[
	\frac{\delta \mathcal H}{\delta s}
	=
	2gs - \frac g{2H} (3s + r)(r - s)
	.
\]
Hence for the Hamiltonian functional \eqref{RS_Hamiltonian}
the bi-directional Whitham system has the form
\begin{align}
\label{RS_sys1}
	r_t &=
	- g \mathcal W r_x
	- \frac g{4H} \mathcal W \partial_x
	(3r + s)(r - s)
	, \\
\label{RS_sys2}
	s_t &=
	g \mathcal W s_x
	- \frac g{4H} \mathcal W \partial_x
	(3s + r)(r - s)
	.
\end{align}
This system is also new even though it has implicitly
appeared in a recently submitted paper \cite{Dinvay_Parau},
where it was not investigated further.
Here we emphasize its usefulness and demonstrate
that this system also outperforms
the system \eqref{Hur_sys1}-\eqref{Hur_sys2} in the sense
of numerical stability.
Moreover, the variables $r$, $s$ have a clear physical meaning
and in particular initial data are easier to obtain.
This means that sometimes the initial elevations
$r(x, 0)$ and $s(x, 0)$ can be measured directly
as opposed to velocity variables.
We do not know if the system is well-posed.
It deserves note that the symbol of
the unbounded operator
$\mathcal W \partial_x$ behaves like a square root
at infinity. This fact might be enough to obtain
well posedness. In any case, as shown later,
the system has favorable numerical stability
properties.
\subsection{Uncoupled twin-unidirectional model.}
\begin{figure}
	\centering
	\subfigure
	{
		\includegraphics[width=0.47\textwidth]
		{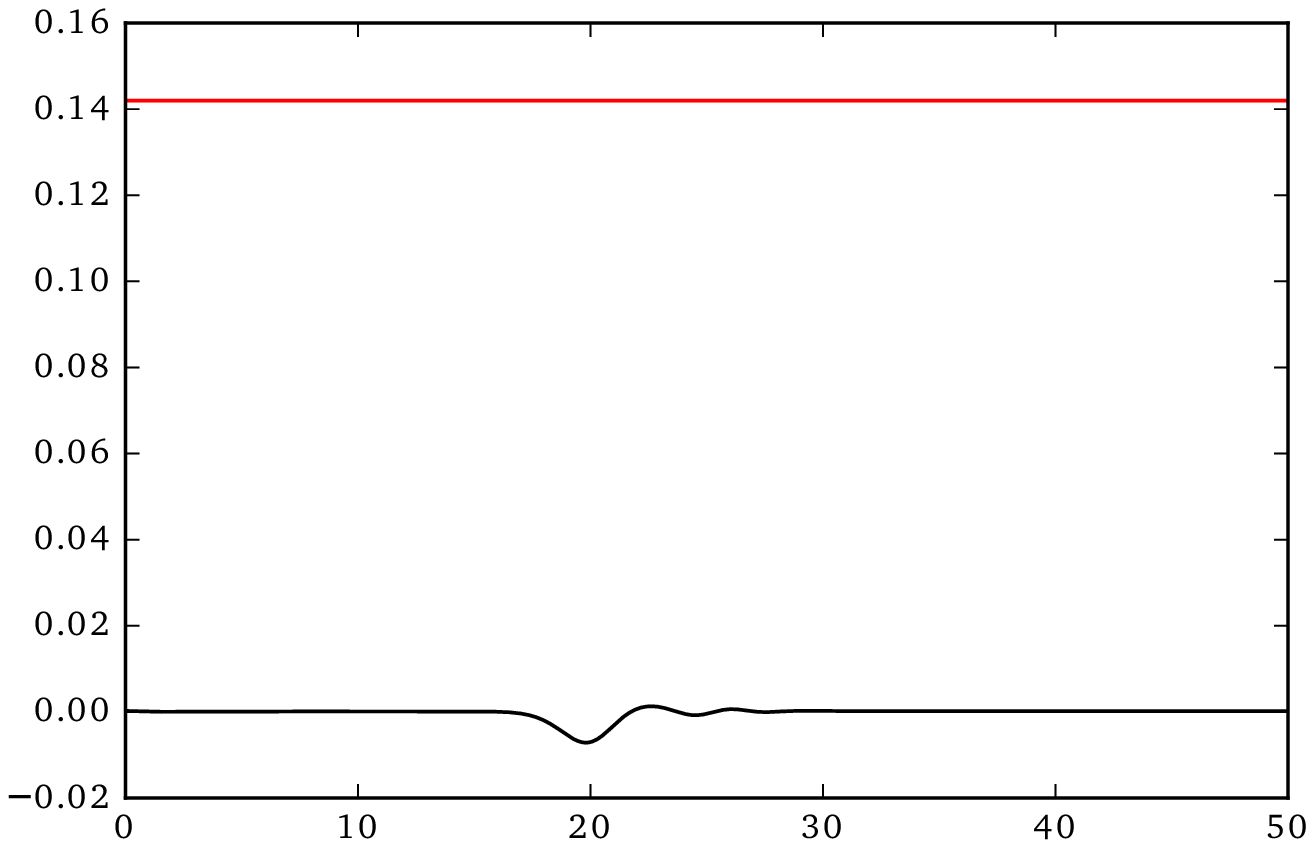}
	}
	\subfigure
	{
		\includegraphics[width=0.47\textwidth]
		{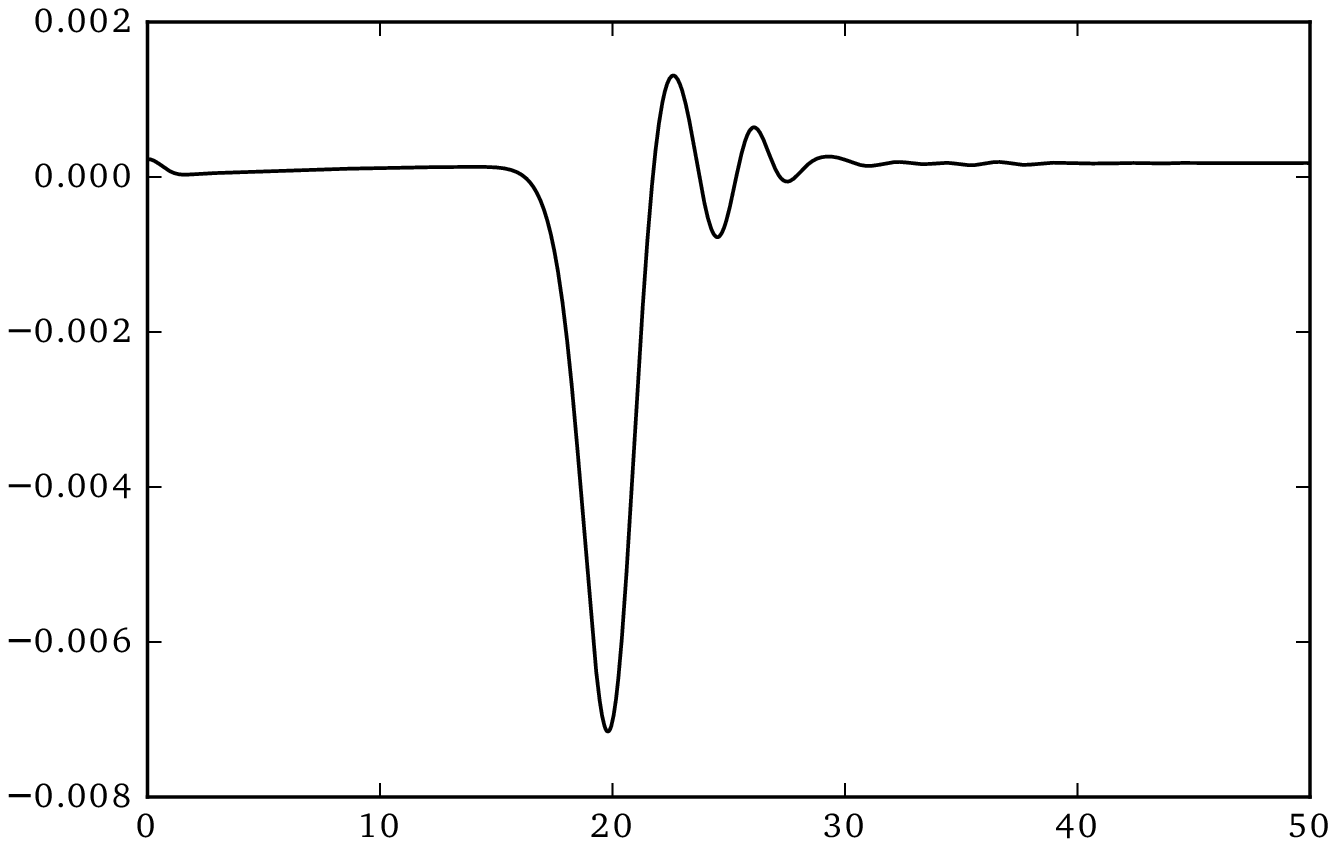}
	}
	\caption
	{\small\em 
	Left panel: Development of the Hamiltonian for total initial energy $\mathcal H = 0.1420$,
        and the coupling term $\mathcal H_{\text{coupling}}$ for Experiment (A).
        Right panel: close-up of the graph of 	$\mathcal H_{\text{coupling}}$.
	}
\label{RScoupling_only}
\end{figure}

One may notice that in the system \eqref{RS_sys1}-\eqref{RS_sys2},
the coupling between the dependent variables is due to
the following part of Hamiltonian \eqref{RS_Hamiltonian}:
\begin{equation}
\label{coupling_Hamiltonian}
	\mathcal H_{\text{coupling}}
	=
	- \frac g{2H} \int_\R rs (r + s) dx.
\end{equation}
This part may sometimes be neglected.
Then we arrive to the Hamiltonian
\begin{equation}
\label{Hamiltonian_without_coupling}
	\mathcal H  = g \int_\R
	\left(	
		r^2 + s^2 + \frac 1{2H} (r^3 + s^3)
	\right)
	dx
\end{equation}
and the corresponding Hamiltonian system consisting of
the two independent equations
\begin{align}
\label{Whitham_sys1_without_coupling}
	r_t &=
	- g \mathcal W r_x
	- \frac {3g}{2H} \mathcal W rr_x
	, \\
\label{Whitham_sys2_without_coupling}
	s_t &=
	g \mathcal W s_x
	+ \frac {3g}{2H} \mathcal W ss_x
	.
\end{align}
The first equation is a modification of the equation
proposed by Whitham \cite{Wh1, Wh2}.
The second one is its analogue for left-going waves.
It is not known if they are well-posed even
though for a large class of similar equations
the answer is affirmative \cite{Ehrnstrom_Pei}.
We shall see below that it is quite often the case
that colliding waves almost do not affect each other
and one may admit independence and regard basically
just the equation \eqref{Whitham_sys1_without_coupling}.
Up to small terms,
the final result is obtained by linear superposition
\eqref{inverse_W_variable_transformation}.
Indeed in Figure \ref{RScoupling_only} the dependence on time
of interaction energy \eqref{coupling_Hamiltonian}
for the Right-left system \eqref{RS_sys1}, \eqref{RS_sys1}
is represented.
One can see that the interaction is going on for a short time
and is of negligible order.
This results in a small residual of solution after the interaction.
%

\section{The numerical approach}
\setcounter{equation}{0}
%
%
All the models discussed in the project are solved by
treating the linear part $\mathcal L$ and the nonlinear part $\mathcal N$
separately using a split-step scheme.
In other words we solve a system of the form
\begin{equation}
\label{general_wave_system}
	z_t = \mathcal L(z) + \mathcal N(z)
\end{equation}
which is treated by solving the systems
\(
	z_t = \mathcal L(z)
\)
and
\(
	z_t = \mathcal N(z)
\).
Denote by $\exp(t \mathcal L)$ an integrator of the first one
and $\exp(t \mathcal N)$ an integrator of the the second one.
We make use of a symplectic integrator of 6th order
introduced by Yoshida \cite{Yoshida}.
The main advantage of such an integrator is that the time step
can be made relatively large
which can accelerate calculations greatly.
Yoshida developed his numerical scheme for
separable finite Hamiltonian systems, however,
it proved to be efficient also in water wave problems \cite{Carter}.
Below we describe the method in application to the models derived.
Following Yoshida a one step integrator for the whole system
\eqref{general_wave_system} is approximated by the product
\[
	\exp [ \delta t (\mathcal L + \mathcal N) )]
	=
	\exp ( c_1 \delta t \mathcal L )
	\exp ( d_1 \delta t \mathcal N )
	\exp ( c_2 \delta t \mathcal L )
	\cdot \ldots \cdot
	\exp ( d_7 \delta t \mathcal N )
	\exp ( c_8 \delta t \mathcal L )
\]
where $\delta t$ is the time step and $c_i$, $d_i$ are constants
given by
\[
	c_1 = c_8 = w_3 / 2
	, \quad
	c_2 = c_7 = (w_3 + w_2) / 2
	, \quad
	c_3 = c_6 = (w_2 + w_1) / 2
	, \quad
	c_4 = c_5 = (w_1 + w_0) / 2
\]
and
\[
	d_1 = d_7 = w_3
	, \quad
	d_2 = d_6 = w_2
	, \quad
	d_3 = d_5 = w_1
	, \quad
	d_4 = w_0
	.
\]
Here we take the following set of weights
\[
    w_3 = 0.784513610477560
	, \quad
	w_2 = 0.235573213359357
    ,
\]
\[
	w_1 = -1.17767998417887
	, \quad
    w_0 = 1.315186320683906
	.
\]
One can notice that the integrator is symmetric.
The meaning of the product is that each time step
is divided into substeps.

The systems
\(
	z_t = \mathcal L(z)
\)
and
\(
	z_t = \mathcal N(z)
\)
are solved using spectral methods.
Moreover, the first one for each model can be solved exactly.
For example, the linearization of the system
\eqref{Hamiltonian_Hur_sys1}-\eqref{Hamiltonian_Hur_sys2}
has the following solution
\[
	\eta(t) =
	\cos Ut \eta_0 - iHD \frac{ \sin Ut }{ U } v_0
	,
\]
\[
	v(t) =
	-ig / H \tanh HD \frac{ \sin Ut }{ U } \eta_0 + \cos Ut v_0,
\]
with the initial data $\eta_0$, $v_0$.
The operator $U$ has the form
\begin{equation}
\label{U_operator}
	U = \sqrt{ gG_0 } = \sqrt{ g D \tanh HD } 
	.
\end{equation}
These formulas represent the integrator $\exp(t \mathcal L)$
for the systems
\eqref{Hamiltonian_Hur_sys1}-\eqref{Hamiltonian_Hur_sys2}
and
\eqref{Hur_sys1}-\eqref{Hur_sys2}
since the linear part $\mathcal L$ is the same for those two.

For the systems \eqref{ASMP_sys1}-\eqref{ASMP_sys2}
and
\eqref{Matsuno_sys1}-\eqref{Matsuno_sys2}
the integrator $\exp(t \mathcal L)$
has the form
\[
	\eta(t) =
	\cos Ut \eta_0 - i\tanh HD \frac{ \sin Ut }{ U } u_0
	,
\]
\[
	u(t) =
	-igD \frac{ \sin Ut }{ U } \eta_0 + \cos Ut u_0,
\]
with the initial data $\eta_0$, $u_0$.

For the system \eqref{RS_sys1}-\eqref{RS_sys2}, 
the operator $\exp(t \mathcal L)$ is diagonal,
\[
	g \mathcal W \partial_x = i g \mathcal W D
	= i \sqrt{ gG_0 } \sgn D,
\]
and the linearized problem has the solution
\[
	r(t) =
	\exp(-itU \sgn D) r_0
	,
\]
\[
	s(t) =
	\exp(itU \sgn D) s_0,
\]
where $r_0$, $s_0$ are initial right- and left-going waves, respectively,
and $U$ is defined by \eqref{U_operator}.

For all models discussed here, we use
the standard Runge-Kutta scheme of 4th order
as the nonlinear integrator $\exp(t \mathcal N)$.
It is explicit but not symplectic.
One might argue that it makes the whole integrator
$\exp( t\mathcal L + t\mathcal N )$ not symplectic any more.

As an alternative we also ran all computations with 
a symplectic Euler scheme, such as described in \cite{Hairer_Wanner_Lubich}.
This scheme turns out to be explicit for most of the models discussed here.
Indeed, for example, for the ASMP model \eqref{ASMP_sys1}-\eqref{ASMP_sys2}
one step of the semi-implicit Euler method has the form
\begin{align*}
	\eta_{n+1} &=
	\eta	_n
	- \delta t \partial_x ( H \mathcal K u_n + \eta_{n+1} u_n )
	, \\
	u_{n+1} &=
	u_n
	- \delta t \partial_x ( g \eta_{n+1} + \frac 12 u_n^2 )
\end{align*}
that can be resolved with respect to $\eta_{n+1}$ as follows.
On the space $l_2^N$ define operator
\(
	B_n f = - \delta t \partial_x (u_nf)
\)
that is bounded
\(
	\lVert B_n \rVert \leqslant \delta t N \max u_n
	.
\)
Expecting uniform boundedness of solution $u$ one can choose
the time step $\delta t = O(1/N)$ so that
\(
	\lVert B_n \rVert \leqslant C < 1
	.
\)
Thus
\[
	(1 - B_n) \eta_{n+1} =
	\eta_n - \delta t \partial_x H \mathcal K u_n
\]
is resolved as
\[
	\eta_{n+1} =
	(1 + B_n + B_n^2 + \ldots)
	( \eta_n - \delta t \partial_x H \mathcal K u_n )
	.
\]
Hence $\eta_{n+1}$, $u_{n+1}$ are resolved via
$\eta_{n}$, $u_{n}$ and the scheme is explicit and
symplectic at the same time.

The numerical scheme of the free-surface problem for the Euler equations
is based on a time-dependent conformal mapping
of the fluid domain into a strip.
A complete description of the method can be found in
\cite{Li_Hyman_Choi, Viotti_Dutykh_Dias}.

%
%
\section{Numerical experiments}
\setcounter{equation}{0}
%
%
The model systems described above are now characterized with respect 
to numerical instability due to spatial discretization.
For the numerical experiments we make the problem nondimensional
by setting $H = 1$ and $g = 1$.
The computational domain is $-L\leq x \leq L$, with $L = 70$.
Initial conditions are imposed by means of
\begin{equation}
\label{test_1}
	\eta_0(x; x_0, a, \lambda) = a\cdot \sech^2(f(x - x_0))-C, 
\end{equation}
where 
\[
	f(\lambda) = \frac{2}{ \lambda }
	\log \left( 1 + \sqrt{2} \right)
	, \qquad
	C(\lambda)
	=
	\frac a{2fL} \left( \tanh f(L - x_0) + \tanh f(L + x_0) \right)
	.
\]
Here $C(\lambda)$ and $f(\lambda)$ are chosen so that 
$\int_{-L}^{L}\eta_0(x)dx = 0$, and
the wave-length $\lambda$
is the distance between the two points $x_1$ and $x_2$
at which $\eta_0(x_1) = \eta_0(x_2) = a/2$.
Below we always take the wave-length $\lambda = \sqrt{5}$.

In all problems below we are interested in time evolution
from $t_0 = 0$ to $t_{max} = 50$.
In cases of collision of two waves we send them towards each other.
So first of all we simulate problems that cannot be described
by unidirectional models like KdV or Whitham equations.
Secondly, one can see that all the models introduced are
in  line with the effect of quasi-elastic interaction of waves.
So after collision waves behave as independent with slight tails.
In all experiments below we provide initial data $\eta(x, 0)$
and $\Phi(x, 0)$ for the Euler system.
Initial data for the approximate models can easily be obtained
by applying transformations of variables
$u(x, 0) = \partial_x \Phi(x, 0)$, \eqref{Hur_transformation}
and \eqref{W_variable_transformation}.
According to \eqref{inverse_W_variable_transformation}
one can make quasi-right moving waves
taking the surface velocity $u(x, 0) = \mathcal W^{-1} \eta(x, 0)$.

As was already said the splitting method we are making use of
allows us to take relatively large time steps.
So we take $\delta t = 0.05$ when the number of Fourier harmonics
is either $N = 512$ or $N = 1024$.
This choice is dictated by the stiffness of the ASMP model
\eqref{ASMP_sys1}-\eqref{ASMP_sys2} since
the scheme becomes unstable for large $N$ and might need filtering
due to the probable ill-posedness of the model.
In comparative experiments, on the other hand, we do not want to use
any filtration.
\begin{experiment}[A]
	Consider a collision of two approaching positive waves.
	Let $a=0.2$ and $x_0=20$.
	Impose initial surface
	\[
		\eta(x, 0) = \eta_0(x;x_0) + \eta_0(x;-x_0)
	\]
	and initial potential
	\[
		\Phi(x, 0)
		=
		- \int_0^x \mathcal W^{-1} \eta_0 (\xi; x_0) d\xi
		+ \int_0^x \mathcal W^{-1} \eta_0 (\xi; -x_0) d\xi
		.
	\]
\end{experiment}
All approximate systems in Experiment (A) are solved on the grid with
$N = 1024$.
\begin{experiment}[B]
	Consider a collision of a trough and a convex wave.
	Let $a=0.1$ and $x_0=20$.
	Impose initial surface
	\[
		\eta(x, 0) = \eta_0(x;x_0) - \eta_0(x;-x_0)
	\]
	and initial potential
	\[
		\Phi(x, 0)
		=
		- \int_0^x \mathcal W^{-1} \eta_0 (\xi; x_0) d\xi
		- \int_0^x \mathcal W^{-1} \eta_0 (\xi; -x_0) d\xi
		.
	\]
\end{experiment}
All approximate systems in Experiment (B) are solved on the grid with
$N = 512$.
\begin{experiment}[C]
	Consider a collision of two troughs.
	Let $a=0.1$ and $x_0=20$.
	Impose initial surface
	\[
		\eta(x, 0) = - \eta_0(x;x_0) - \eta_0(x;-x_0)
	\]
	and initial potential
	\[
		\Phi(x, 0)
		=
		\int_0^x \mathcal W^{-1} \eta_0 (\xi; x_0) d\xi
		- \int_0^x \mathcal W^{-1} \eta_0 (\xi; -x_0) d\xi
		.
	\]
\end{experiment}
All approximate systems in Experiment (C) are solved on the grid with
$N = 512$.
\begin{experiment}[E1-E3]
	Consider the evolution of waves with the initial surface elevation
	\[
		\eta(x, 0) = \eta_0(x; x_0 = 0)
	\]
	where $a=0.3$ and $x_0=0$.
	Impose firstly (E1) initial potential
	\[
		\Phi(x, 0)
		=
		\int_0^x \mathcal W^{-1} \eta_0 (\xi) d\xi
		,
	\]
	than secondly (E2) initial potential
	\[
		\Phi(x, 0)
		=
		\int_0^x \eta_0 (\xi) d\xi
		,
	\]
	and finely (E3) initial potential
	\[
		\Phi(x, 0) = 0
		.
	\]
\end{experiment}
All approximate systems in Experiments (E1-E3) are solved on the grid with
$N = 1024$.
Note that the initial potential of Experiment (E2) creates only
approximately a right-going wave according to the linear long wave theory.
Anyway neither the conditions of Experiment (E1) or of Experiment (E2)
induce completely one way propagation as numerical results shows.
Surprisingly, initial potentials of the type as in Experiment (E3)
lead to better correspondence between approximate models
and the Euler system then
initial potentials of the type as in Experiment (E2).
And moreover, of the type as in Experiment (E2)
lead to the better correspondence then of the type as in Experiment (E1).
We believe it is mainly a technical feature since the initial error
of evaluation surface potential via $\mathcal W^{-1}$ and integration
normally increases with the time.
\begin{figure}
	\centering
	\subfigure
	{
		\includegraphics[width=0.99\textwidth]
		{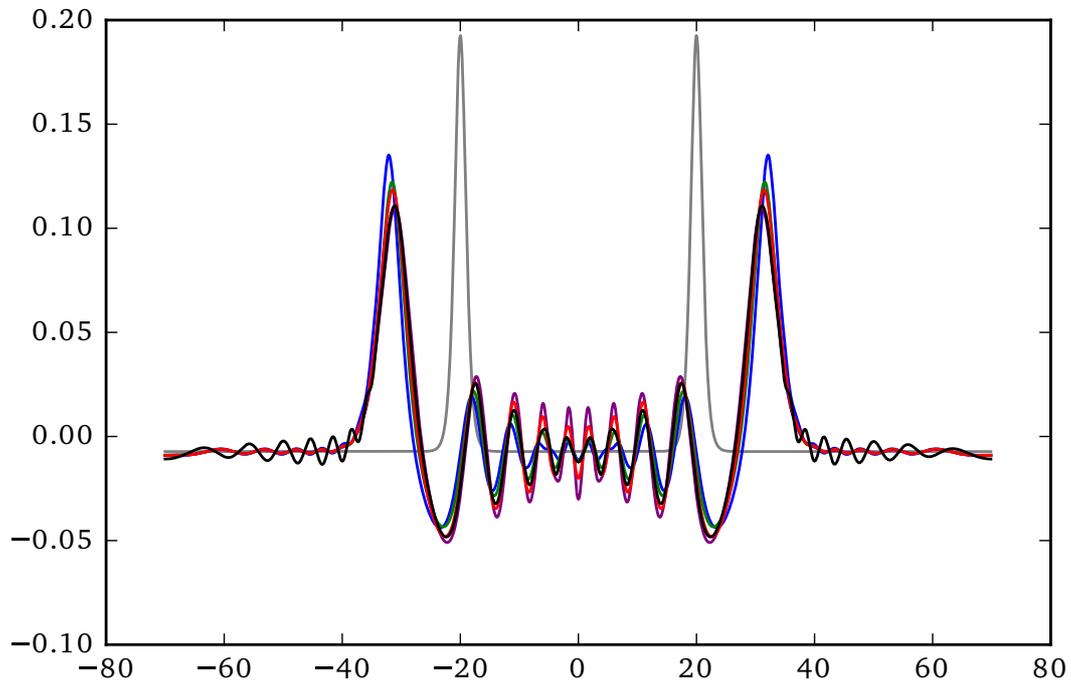}
	}
	\caption
	{\small\em Experiment (A). The thin grey curve represents the initial data. The black curve is the approximate
         solution of the full Euler system at $t=50$. The color coding is as follows: 
         purple -- Hamiltonian HP system; red -- right-left system; 
         blue -- ASMP system; green -- HP system.
	}
\label{figureA}
\end{figure}
\begin{figure}
	\centering
	\subfigure
	{
		\includegraphics[width=0.99\textwidth]
		{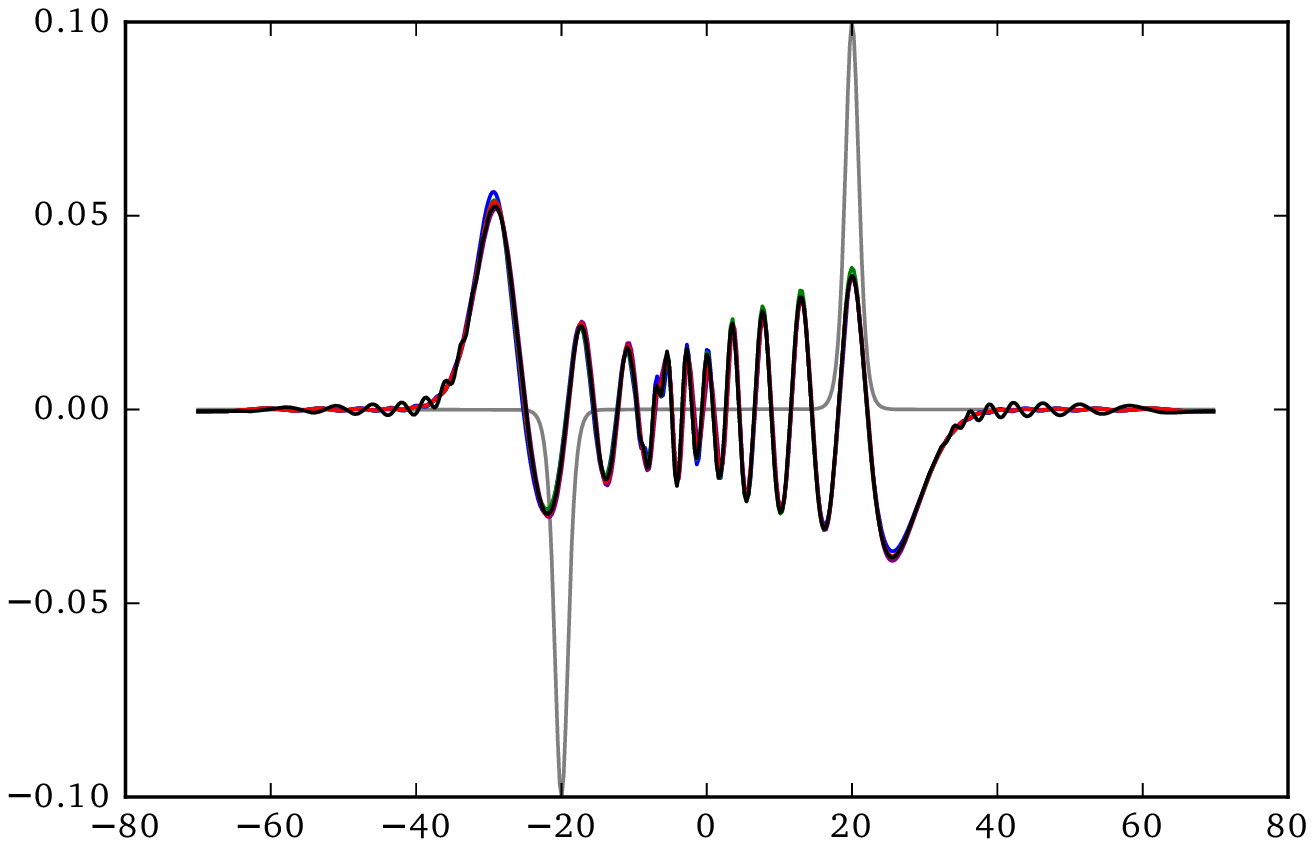}
	}
	\caption
	{\small\em Experiment (B). The thin grey curve represents the initial data. The black curve is the approximate
         solution of the full Euler system at $t=50$. The color coding is as follows: 
         purple -- Hamiltonian HP system; red -- right-left system; 
         blue -- ASMP system; green -- HP system.
	}
\label{figureB}
\end{figure}
\begin{figure}
	\centering
	\subfigure
	{
		\includegraphics[width=0.99\textwidth]
		{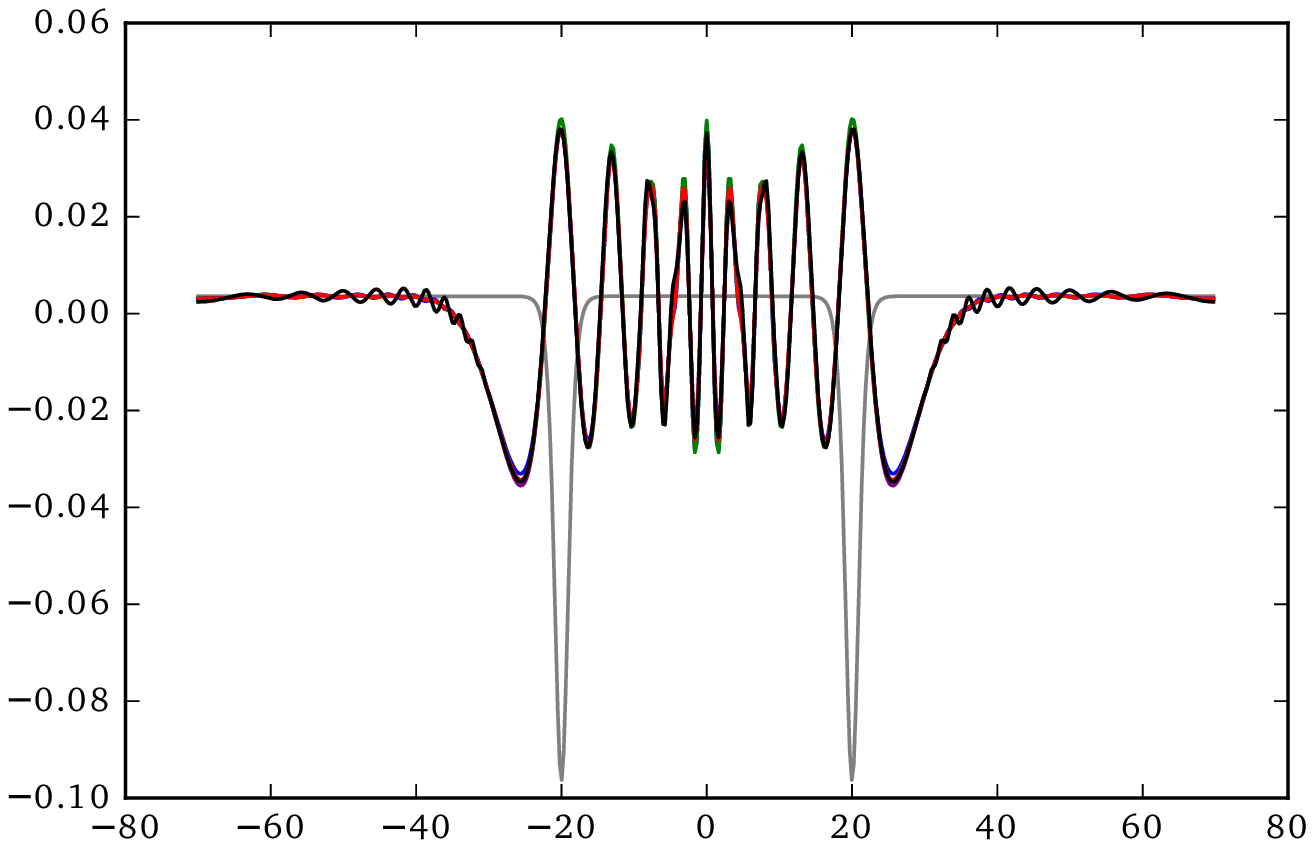}
	}
	\caption
	{\small\em 
         Experiment (C). The thin grey curve represents the initial data. The black curve is the approximate
         solution of the full Euler system at $t=50$. The color coding is as follows: 
         purple -- Hamiltonian HP system; red -- right-left system; 
         blue -- ASMP system; green -- HP system.
	}
\label{figureC}
\end{figure}
\begin{figure}
	\centering
	\subfigure
	{
		\includegraphics[width=0.99\textwidth]
		{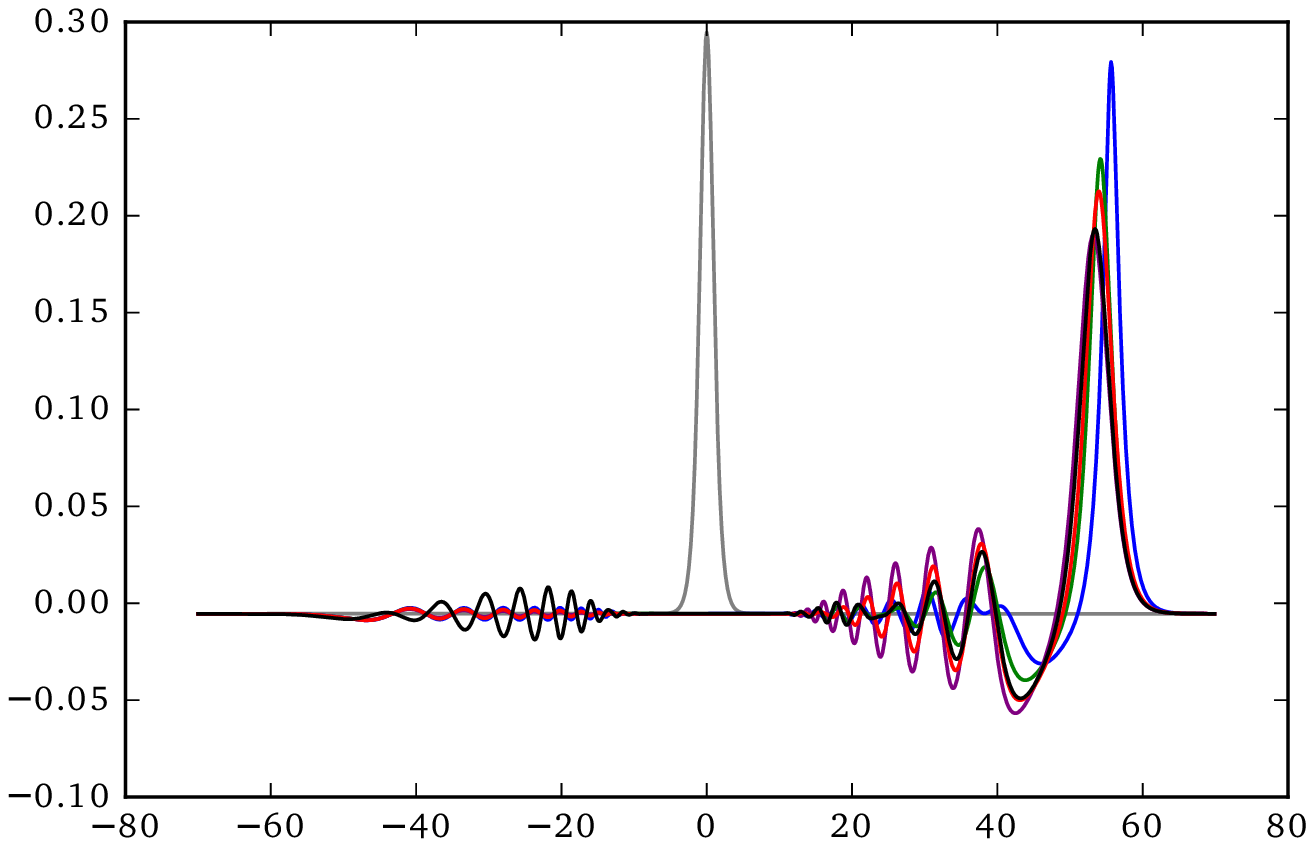}
	}
	\caption
	{
		Experiment (E1). The thin grey curve represents the initial data. The black curve is the approximate
         solution of the full Euler system at $t=50$. The color coding is as follows: 
         purple -- Hamiltonian HP system; red -- right-left system; 
         blue -- ASMP system; green -- HP system.
	}
\label{figureE1}
\end{figure}
\begin{figure}
	\centering
	\subfigure
	{
		\includegraphics[width=0.99\textwidth]
		{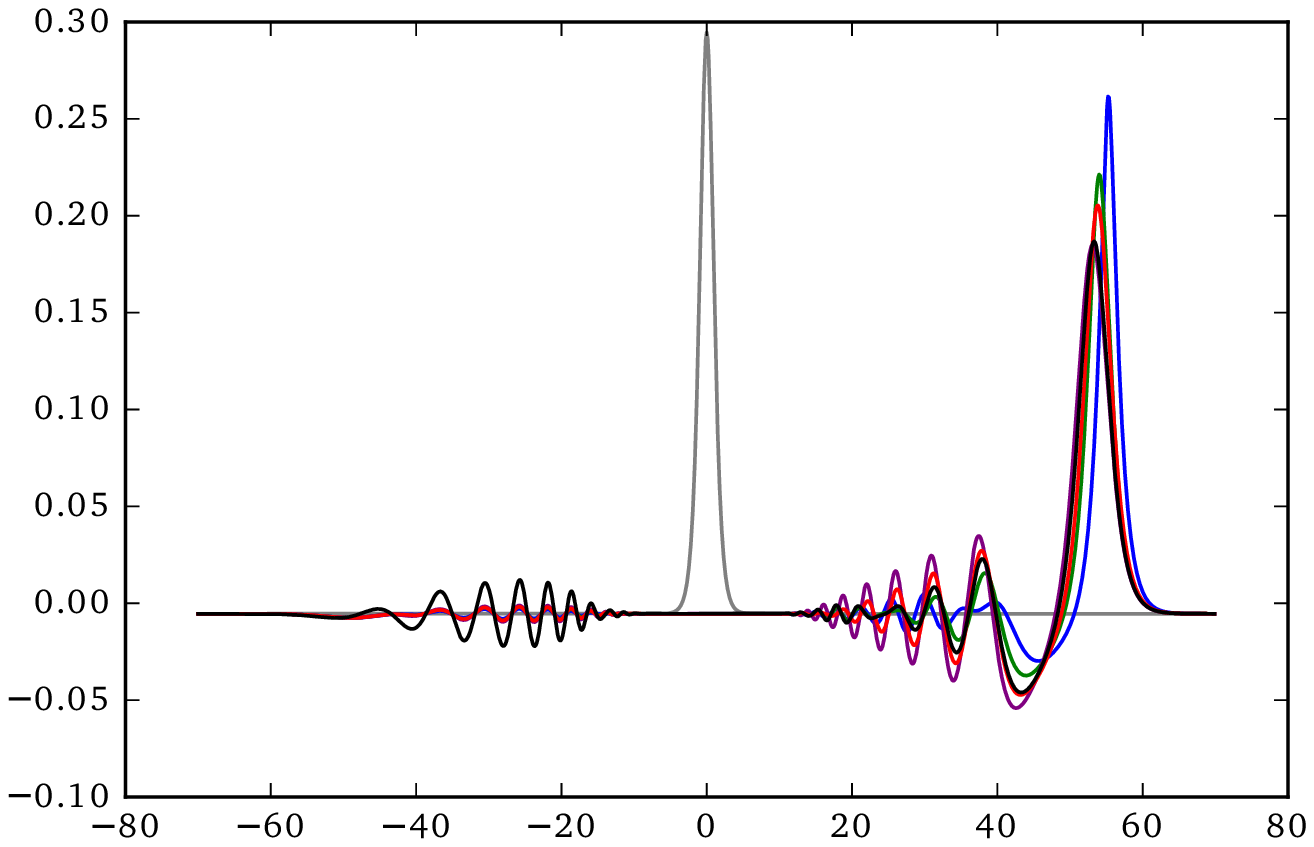}
	}
	\caption
	{\small\em 
		Experiment (E2). The thin grey curve represents the initial data. The black curve is the approximate
         solution of the full Euler system at $t=50$. The color coding is as follows: 
         purple -- Hamiltonian HP system; red -- right-left system; 
         blue -- ASMP system; green -- HP system.
	}
\label{figureE2}
\end{figure}
\begin{figure}
	\centering
	\subfigure
	{
		\includegraphics[width=0.99\textwidth]
		{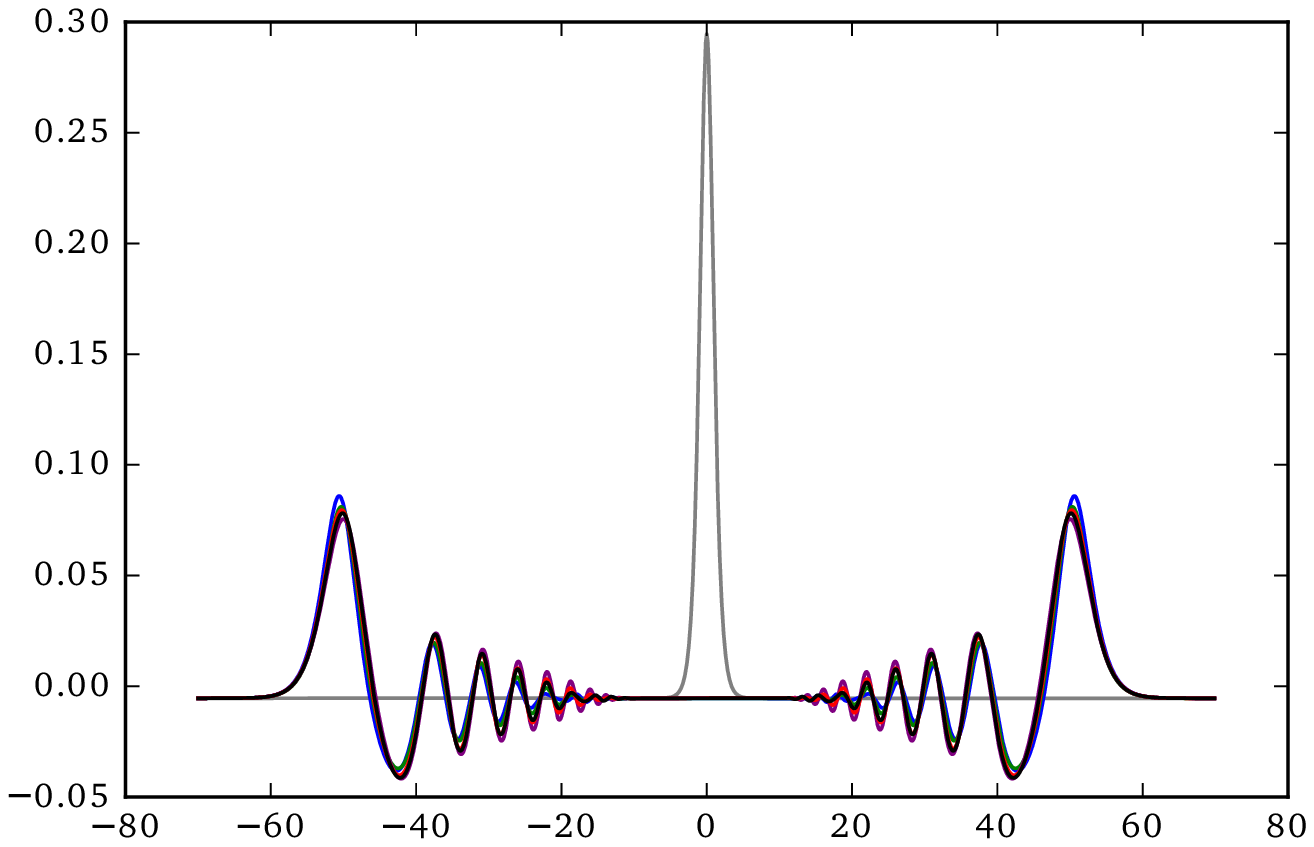}
	}
	\caption
	{\small\em 
		Experiment (E3). The thin grey curve represents the initial data. The black curve is the approximate
         solution of the full Euler system at $t=50$. The color coding is as follows: 
         purple -- Hamiltonian HP system; red -- right-left system; 
         blue -- ASMP system; green -- HP system.
	}
\label{figureE3}
\end{figure}

In all presented figures initial elevation profiles are marked by grey lines.
Solutions of the Euler system \eqref{Euler_sys1}-\eqref{Euler_sys4}
are black,
of the ASMP system \eqref{ASMP_sys1}-\eqref{ASMP_sys2} are blue,
of the Hur--Pandey system \eqref{Hur_sys1}-\eqref{Hur_sys2} are green,
of the Hamiltonian Hur--Pandey system
\eqref{Hamiltonian_Hur_sys1}-\eqref{Hamiltonian_Hur_sys2} are purple,
and of the right-left system \eqref{RS_sys1}-\eqref{RS_sys2} are red.

\begin{table}
\begin{center}
	\begin{tabular}{| c | c | c | c | c | c | c |}
		\hline
		Experiment  &  A  &  B  &  C  &  E1  & E2 & E3 
		\\
		\hline
		Euler	&  0.1316  &  0.0329075955585  &  0.03291  &
		0.1481  & 0.1398 & 0.0740610317118
		\\	
		ASMP 	&  0.1440  &  0.0329075170851  &  0.03136  & 
		0.1686  & 0.1569 & 0.0740419134333
		\\
		Hamiltonian HP	&  0.1405  &  0.0329075170854  &  0.03180  & 
		0.1626  & 0.1524 & 0.0740419134422
		\\
		Right--Left	&  0.1420  &  0.0329075170854  &	0.03162  & 
		0.1651  & 0.1543 & 0.0740419134422
		\\
		\hline
	\end{tabular}
\end{center}
\bigskip
\caption{Hamiltonians $\mathcal H$ for various systems, evaluated at $t=50$.}
\end{table}

\begin{table}
\begin{center}
	\begin{tabular}{| c | c | c | c | c | c | c |}
		\hline
		Experiment  &  A  &  B  &  C  &  E1  & E2 & E3 
		\\
		\hline
		ASMP 	&  0.488  &  0.109  &  0.149  &
		0.883  & 0.768 & 0.153
		\\	
		Hur--Pandey  	&  0.253  &  0.085  &  0.126  & 
		0.339  & 0.315 & 0.082
		\\
		Hamiltonian HP	&  0.167  &  0.130  &  0.106  & 
		0.231  & 0.207 & 0.061
		\\
		Right--Left	&  0.167  &  0.089  &	0.128  & 
		0.240  & 0.218 & 0.048
		\\
		\hline
	\end{tabular}
\end{center}
\bigskip
\caption{Errors $\mathcal E$, evaluated at $t=50$.}
\end{table}

In order to quantitatively compare the accuracy of each approximate model
we calculate the differences between Euler solutions
and solutions of each system correspondingly.
These errors are measured in the integral $L^2$-norm
normalized by initial condition as follows
\[
	\mathcal E
	=
	\frac{ \lVert \eta_{Euler} - \eta_{model} \rVert }
	{ \lVert \eta_{initial} \rVert }
\]
where
\[
	\lVert \eta_{Euler} - \eta_{model} \rVert
	=
	\max_{t}
	\sqrt
	{
		\int ( \eta_{Euler}(x, t) - \eta_{model}(x, t) )^2 dx
	}
\]
and
\[
	\lVert \eta_{initial} \rVert
	=
	\sqrt
	{
		\int \eta(x, 0)^2 dx
	}
	.
\]
Here $ \eta_{Euler}(x, t) $ is the solution for the Euler system
and $ \eta_{model}(x, t) $ corresponds either to
ASMP, Hur--Pandey, Hamiltonian Hur--Pandey or Right--Left system.
The corresponding results are represented in Table 2.

\begin{figure}
	\centering
	\subfigure
	{
		\includegraphics[width=0.99\textwidth]
		{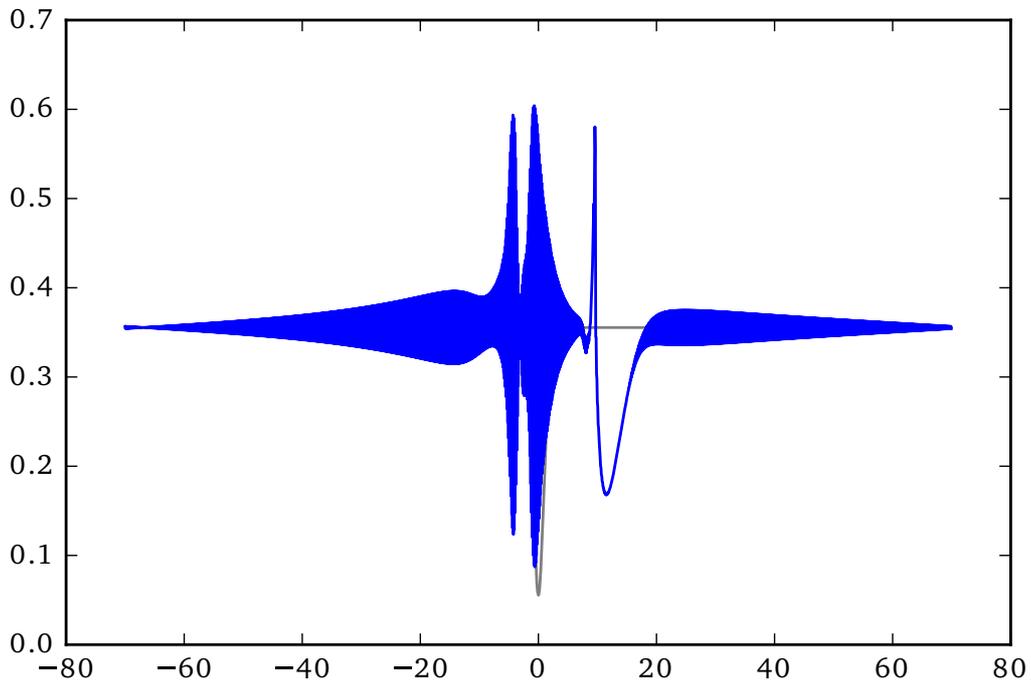}
	}
	\caption
	{\small\em 
		Approximate solution of the ASMP Whitham system with initial data 
                satisfying the condition $\inf{\eta_0} > 0$. 
	}
\label{ASMP_ill_posed_test}
\end{figure}

As was stated above some models work better in the sense of
numerical stability.
There were many discussions about ill-posedness of ASMP model
\cite{Claassen_Johnson}.
In the next experiment we provide an example
with initial data satisfying the condition for local well posedness.
One can see that the initial data is lifted over
the real axis so the mean value is approximately 0.35.
It is known from Ehrnstr\"om, Pei, Wang \cite{Pei_Wang}
that we are in a locally well posed situation, 
however, the obtained solution seems very unstable
as one can see in Figure \ref{ASMP_ill_posed_test}.
This experiment was repeated with different time integrators,
including the symplectic first-order Euler method described in Section 4. 
The results were always the same, pointing to doubts about the long-time
well posedness of the ASMP system.

In order to systematize our experiments regarding the
well posedness and stability of the Whitham systems,
we used the following initial data:
\begin{experiment}
	Suppose we have a trough with amplitude $a= 0.3$.
	Let $x_0 = 0$.
	Solve System \eqref{ASMP_sys1}-\eqref{ASMP_sys2}
	with the initial surface
	\[
		\eta(x, 0) = - \eta_0(x) + 0.35
	\]
	and the initial velocity
	\[
		u(x, 0)
		=
		\mathcal W^{-1} \eta(x, 0)
		.
	\]
\end{experiment}

\begin{figure}
	\centering
	\subfigure
	{
		\includegraphics[width=0.47\textwidth]
		{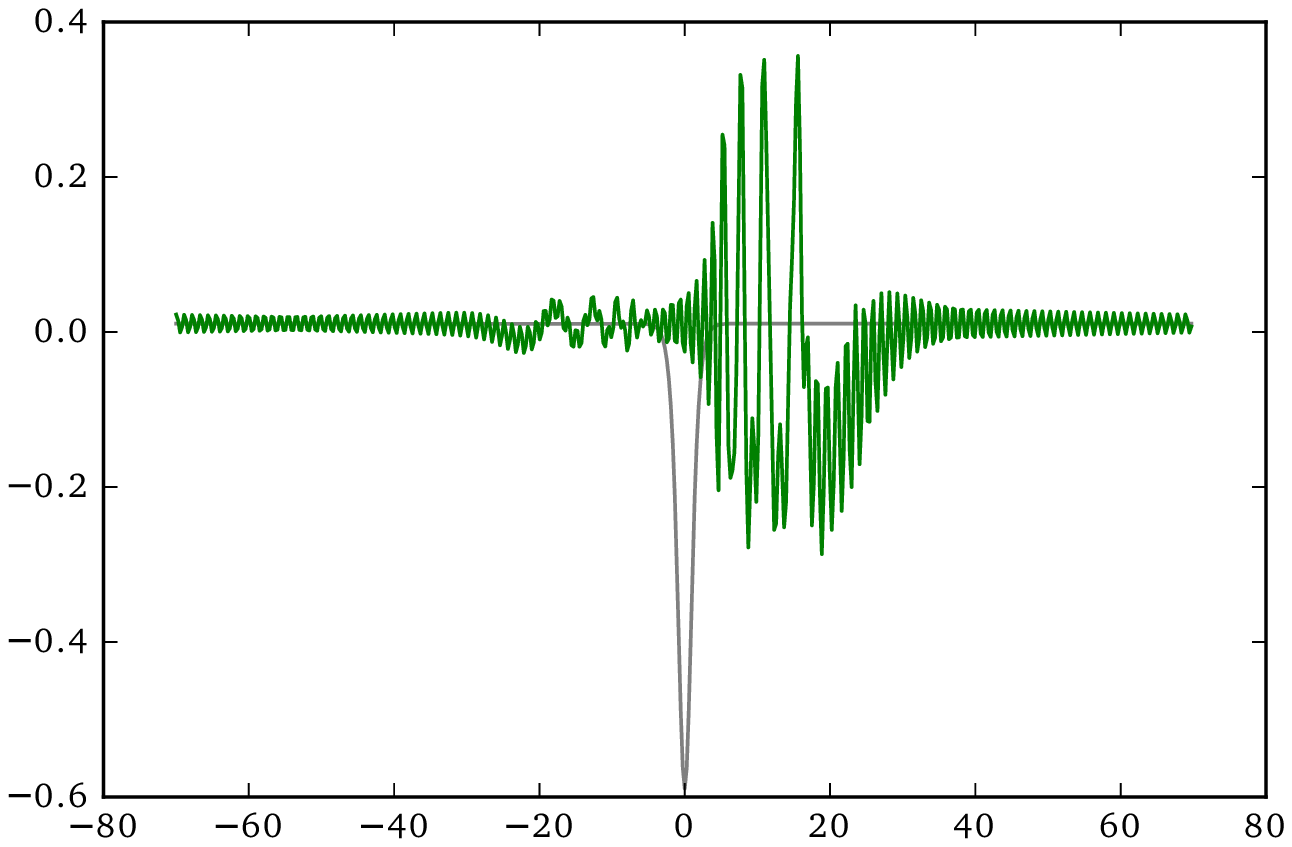}
	}
	\subfigure
	{
		\includegraphics[width=0.47\textwidth]
		{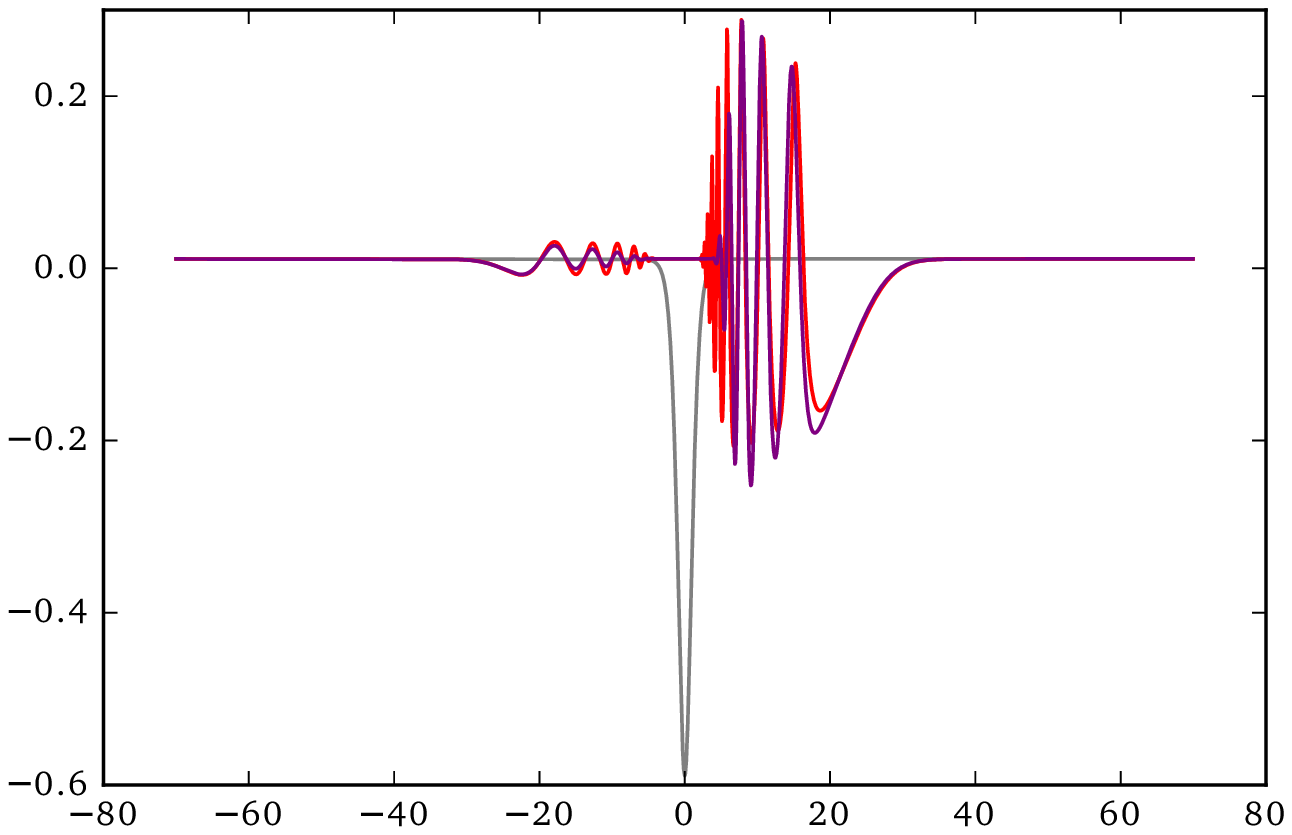}
	}
	\caption
	{\small\em 
		Left panel: De-aliased solution of the HP system
		with N = 512 and time step = 0.05.
		Snapshot is taken at $t = 25$.
		Right panel: The same for Hamiltonian version of the HP system and the Right--Left system.
	}
\label{Hur_alising}
\end{figure}

Problems with the HP system \eqref{Hur_sys1}-\eqref{Hur_sys2}
may occur if an initial trough is deep enough.
In the example shown on Figure \ref{Hur_alising} we have to filter half
of the high Fourier modes to make computations stable.
The resulting noisy solution continues its propagation
and one can notice that all the oscillations happen around
some reasonable mean curve that can be obtained easily
by solving either the system
\eqref{Hamiltonian_Hur_sys1}-\eqref{Hamiltonian_Hur_sys2}
or the system \eqref{RS_sys1}-\eqref{RS_sys2} without any filtration.
The results are represented on Figure \ref{Hur_alising}.

\begin{experiment}
	Suppose $a= 0.6$ and $x_0 = 0$.
	Solve System \eqref{Hur_sys1}-\eqref{Hur_sys2}
	with initial surface
	\[
		\eta(x, 0) = - \eta_0(x)
	\]
	and initial velocity
	\[
		v(x, 0)
		=
		\mathcal K \mathcal W^{-1} \eta(x, 0)
		.
	\]
\end{experiment}

As to numerical stability of the Right--Left system
\eqref{RS_sys1}-\eqref{RS_sys2}, we can notice that this system
encountered problems only in extreme non-physical situations,
as for example, with an initial deep trough of amplitude
$a = 1.2$ and increasing number of harmonics up to
$N = 2^{15}$.
The Hamiltonian version of the
Hur--Pandey system
\eqref{Hamiltonian_Hur_sys1}--\eqref{Hamiltonian_Hur_sys2}
is numerically stable even in such a physically absurd problem.

Finally, let us look at the development of the Hamiltonian in two cases.
First, an example of self-stabilization in the Matsuno system:
\begin{experiment}
	Suppose $a= 0.2$ and $x_0 = 0$.
	Solve Matsuno System \eqref{Matsuno_sys1}-\eqref{Matsuno_sys2}
	with initial surface
	\(
		\eta(x, 0) = \eta_0(x)
	\)
	and initial velocity
	\(
		u(x, 0)
		=
		\mathcal K \mathcal W^{-1} \eta(x, 0)
		.
	\)
	We take the time step $\delta t = 0.1$ and the number
	of grid points $N = 512$.
\end{experiment}

\begin{figure}
	\centering
	\subfigure
	{
		\includegraphics[width=0.47\textwidth]
		{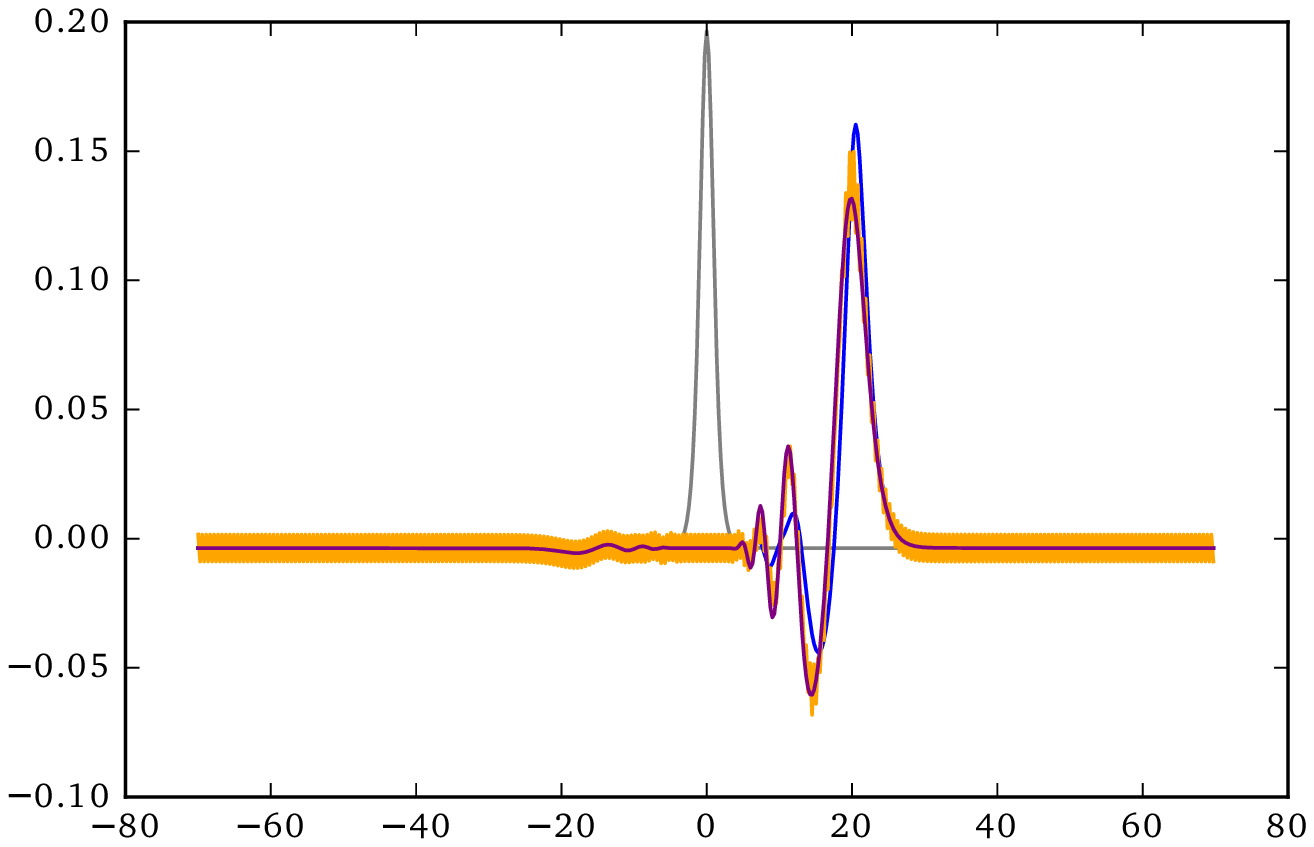}
	}
	\subfigure
	{
		\includegraphics[width=0.47\textwidth]
		{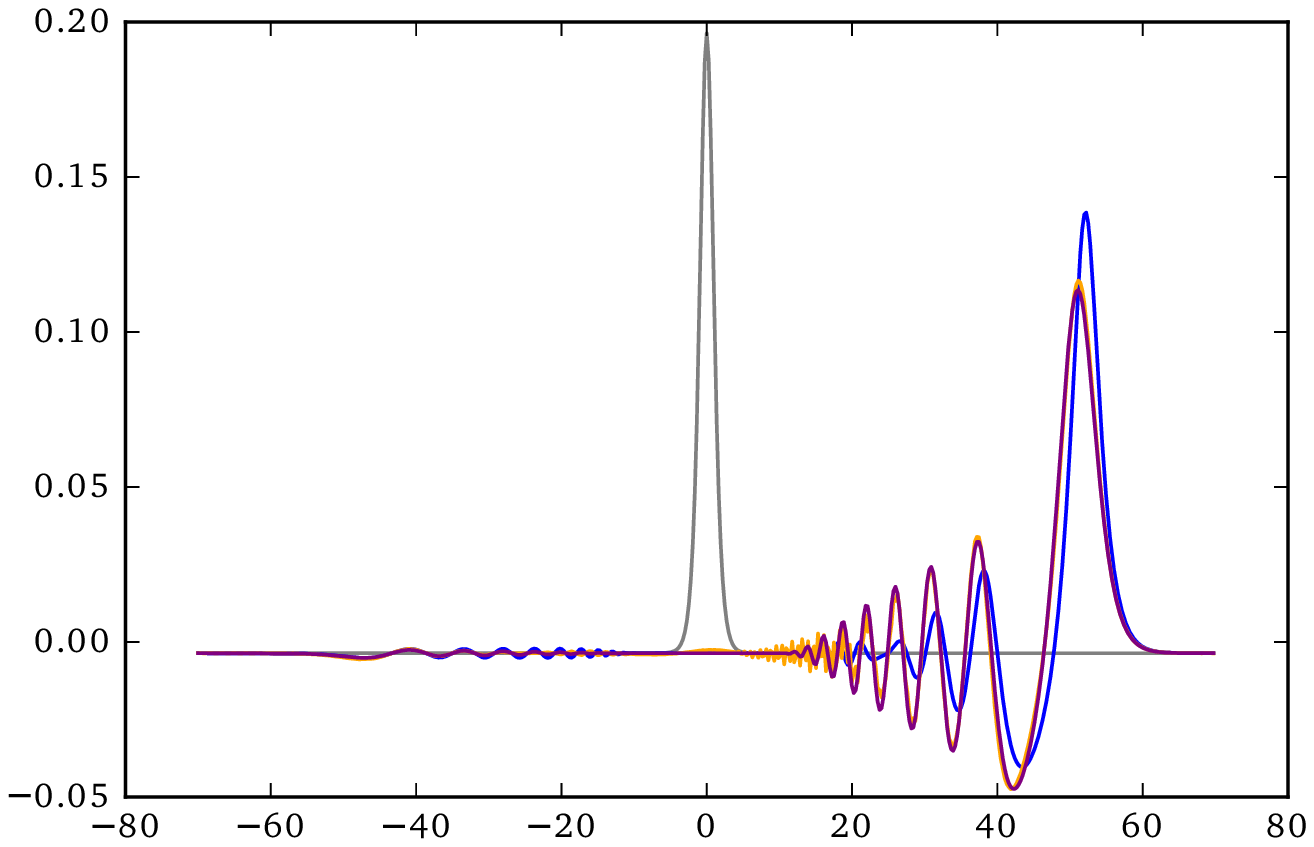}
	}
	\caption
	{\small\em 
		Self-stabilized solution of Matsuno system
		with $N = 512$ and time step $\delta t = 0.1$.
		Left panel: $t = 20$, right panel: $t = 50$.
	}
\label{Matsuno_stabilisation}
\end{figure}
\begin{figure}
	\centering
	\subfigure
	{
		\includegraphics[width=0.99\textwidth]
		{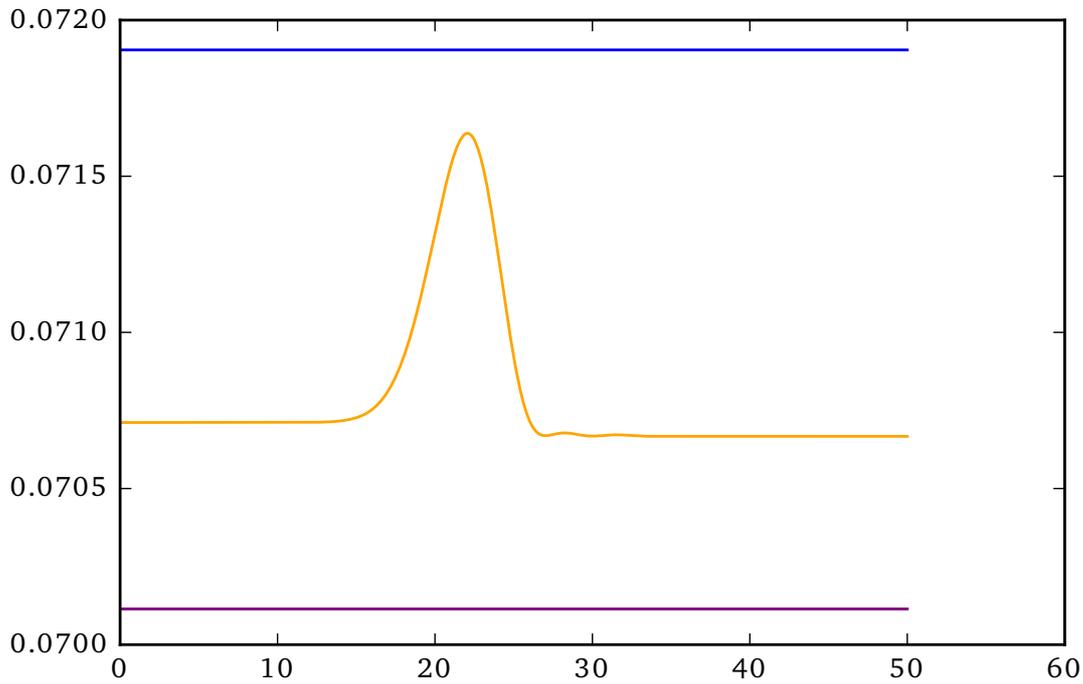}
	}
	\caption
	{\small\em 
		Total energy of the self-stabilized solution of Matsuno system
		as a function of time $t$, with $N = 512$ and time step $\delta t = 0.1$.
	}
\label{Matsuno_stabilisation_Hamiltonian}
\end{figure}
One might think that a numerical method conserving the total energy
could remove the instabilities in the solution.
Unfortunately this is not the case.
We applied a simple projection method \cite{Hairer_Wanner_Lubich}
to obtain a conservative method.
With this method, energy was indeed conserved, and we
managed to get a constant instead of the time-varying energy
shown in Figure \ref{Matsuno_stabilisation_Hamiltonian}.
However, the solutions itself remained noisy 
such as in Figure \ref{Matsuno_stabilisation},
and the computational cost is substantially higher than in the
nonconservative method.

\section{Acknowledgments}
This research was supported in part by the Research Council of Norway through grants 213474/F20 and
239033/F20.

\vspace*{-0.75em}
\end{document}